\documentclass[fleqn,twoside]{article}
\usepackage{espcrc2}
\usepackage{epsfig}
\usepackage{graphicx}

\def\lsi{\raise0.3ex\hbox{$<$\kern-0.75em\raise-1.1ex\hbox{$\sim$}}}
\def\gsi{\raise0.3ex\hbox{$>$\kern-0.75em\raise-1.1ex\hbox{$\sim$}}}

\title{
\vspace*{-1.2cm}
\mbox{} \hfill {\small ITP-Budapest 588, DESY 02-141}\\
\vspace*{0.7cm}
A scalable PC-based parallel computer for lattice QCD}
\author{
Z. Fodor\address[ELTE]{Institute for Theoretical Physics, E\"otv\"os University,
P\'azm\'any P. 1/A, H-1117 Budapest, Hungary}, 
S.D. Katz\address{Deutsches Elektronen-Synchrotron DESY, Notkestr. 85, D-22607,
Hamburg, Germany}\thanks{on leave from
Institute for Theoretical Physics, E\"otv\"os University,
P\'azm\'any P. 1/A, H-1117 Budapest, Hungary} 
and G. Papp\addressmark[ELTE]}

\begin{document}

\begin{abstract}
A PC-based parallel computer
for medium/large scale lattice QCD simulations is suggested. 
The E\"otv\"os Univ., Inst. Theor. Phys. cluster consists
of 137 Intel P4-1.7GHz nodes. Gigabit Ethernet cards
are used for nearest neighbor communication in a two-dimensional mesh.
The sustained performance for dynamical staggered(wilson) quarks
on large lattices is around 70(110) GFlops.
The exceptional price/performance ratio is below \$1/Mflop.
\end{abstract}

\maketitle

\section{Introduction}

The most powerful computers for lattice gauge theory are industrial
supercomputers or special purpose parallel computers (see e.g.
\cite{APE-project,Tsukuba,Columbia}).   
It is more and more accepted that e.g. off-the-shelf  
PC systems can be used to build 
parallel computers for lattice QCD simulations 
\cite{Wuppertal,CSea99,Lea01}.  

Single PC hardware has excellent price/perfor\-mance ratios. 
For recent review papers see \cite{C99,L01}.

We present our experiences and benchmark results on a  
scalable system, which uses nearest-neighbor
communication through Gigabit Ethernet (GigE) cards.
The communication is fast enough (consuming 40\% of the total time
in typical applications) and cheap enough (30\% of the total
price is spent on the communication).
The system can sustain $\approx$100 Gflops on today's medium/large
lattices.  Its price/performance ratio is
below \$1/Mflops for 32-bit applications (twice as much for 
64-bit applications). 

Already in 1999 a report was presented \cite{CSea99}
on the PC-based parallel computer project at the
E\"otv\"os University, Budapest, Hungary. 
A machine was constructed with 32 PCs arranged in a three-dimensional 
2$\times$4$\times$4 mesh. Each node had two special, purpose designed
communication cards providing communication 
to the six neighbors. We used the ``multimedia extension'' instruction set
of the processors to increase the performance.
The communication bandwidth (16 Mbit/s) at that time was only enough for 
bosonic simulations.
Here we report on a system which needs
no hardware development and has two orders of 
magnitude larger communication bandwidth through GigE cards.

\section{Hardware}
Each node consists
of an 
Intel KD850GB motherboard,
Intel-P4-1.7GHz processor,
512~MB RDRAM,
100~Mbit Ethernet card,
20.4~GB IDE HDD and four SMC9452 GigE cards
for the two dimensional inter-node communication. 
The 100~Mbit Ethernet cards with switches are used for job 
management only. 

We have 137 PC nodes. 
A single machine is used for controlling. 
Two smaller clusters with 4 nodes are used  for development.  
The remaining 128 machines can be  used as one cluster with 
128 nodes (or two clusters with 64 nodes or four clusters with 32 nodes) 
for mass production.

The best number of connected nodes 
for a given lattice can be determined by optimizing 
the surface to volume ratio of the local sub-lattice. 
Changing the node topology needs reconnecting 
cables, which can be done easily in a few minutes. 

In April, 2002 the price of one node 
including the 100~Mbit Ethernet 
switches is \$380 (see \cite{pricewatch}). The four GigE cards with cables 
cost additional \$160 for each node. 
The power consumption of the nodes (140W each) requires
a cooling system which costs around \$13 per node.
Thus, the total node (PC+communication+cooling) price is about \$553.

The key element of a cost effective design is an appropriate 
balance between communication and the performance of the nodes.
We spent more than twice as much on the bare PC (\$393
including cooling) than for the Gigabit communication (\$160). 
These numbers are in strong contrast with
Myrinet based PC systems, for which the high price of the Myrinet card
exceeds the price of such a PC by a factor of two. 

\section{Software}

\begin{figure}\begin{center}
\vspace{-1.1cm}
\includegraphics*[width=8.cm]{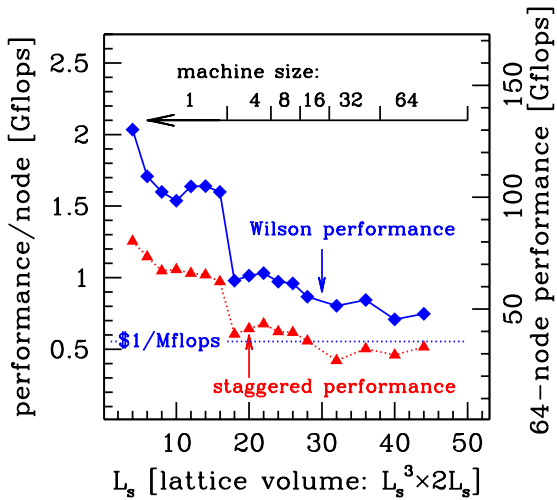}
\vspace{-1.5cm}
\caption{Full QCD performance results on a 64-node cluster.
The node and the total performance 
as a function of the spatial extension $L_s$. 
The lattice sizes are chosen to be $L_s^3\times 2L_s$.
The dotted line represents the \$1/Mflops value. 
Wilson and staggered quarks are both indicated.
The machine size (the number of communicating nodes)
for a given lattice volume is shown by an inserted scale.
\label{perf}}
\vspace{-0.9cm}
\end{center}\end{figure}

The main operating system of the cluster is SuSE Linux 7.1, being
installed on each node. 
Job management is done by the ``main'' computer through the 
100~Mbit Ethernet network. We developed a simple job-management utility
to distribute jobs on the nodes and collect the results.

To take advantage of the Gigabit communication from applications
(e.g. C, C++ or Fortran), a simple C library was written using the
standard Linux network interface. Currently, we are using a standard socket
based communication with Transmission Control Protocol (TCP) widely used
in network applications. The typical bandwidth that we can reach in QCD
applications is around 400~Mbit/s in contrast to the theoretical 1000~Mbit/s.
Writing a lower level driver for the gigabit cards which would 
increase the bandwidth is in progress.
The functions of the library can be 
used to open and close the communcation channels and
transfer data between the neighbors.

\begin{figure}\begin{center}
\includegraphics*[width=6.1cm]{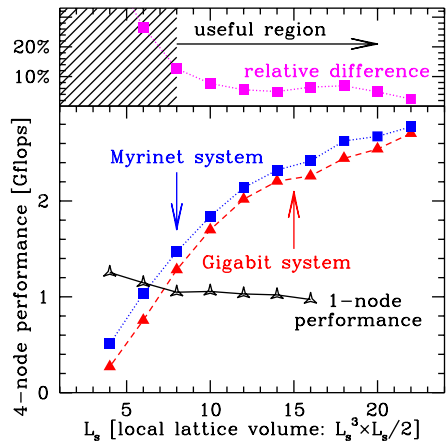}
\vspace{-0.7cm}
\caption{Sustained staggered performance of a 4-node (2$\times$2) system as a
function of the local lattice volume V=$L_s^3\times L_s/2$ (lower panel). 
Squares show
the performance of a myrinet,  
triangles that of  
a gigabit sytem.
The one-node performance is also shown by stars. Communication is of no use
if the one-node performance is better than the 4-node one.
Thus, the useful region is given by local lattices
$8^3\times 4$ or larger. In this useful region the relative difference
between the performances of the myrinet and gigabit systems (upper panel)
is always less than 12\%.
\label{myrinet}}
\vspace{-0.5cm}
\end{center}\end{figure}

\section{Performance}
Figure \ref{perf} gives a summary of our benchmark runs. Single precision
is used for local variables (gauge links and vectors, see \cite{MILC}) 
and dot products are accumulated in double precision.
Using double precision gauge links and vectors reduces the performance by
approximately 50\%. 

For benchmarking we started from the MILC code
~\cite{MILC}.
To increase the performance of the code we modified 
it by three different techniques.

First of all, we used the ``multimedia extension'' 
({\tt sse}) instructions of the 
Intel-P4 processors 
As we pointed out it in 1999 \cite{CSea99} this capability
can accelerate the processor by a large factor.
We rewrote almost the whole conjugate gradient part of the
program to assembly (including also loops over the lattice and not only
elementary matrix operations). 
This way we obtained a speedup factor of $\approx$2. 
(see similar results of e.g. Ref. \cite{L01,P01}).

Secondly, an important speedup was obtained by changing the data structure 
as suggested by S. Gottlieb \cite{G01}. 
In the original ``site major'' concept
all the physical variables of a given site are stored in one structure and the
lattice is an array of these structures.  Instead of this concept one should 
use ``field major'' variables.
The set of a given type of variable of the different
sites are collected and stored sequentially. This increases the number of
cache hits.
Similarly to Ref. \cite{G01} this change leads to
a speedup factor of $\approx$2.

The third ingredient of our improvement was a better cache management by
extensive use of the  
``prefetch'' instruction. 

One of the most obvious features of Figure \ref{perf} is the sharp drop of the
performance when one turns on the communication. The most economic solution
is to turn on the communication only if the memory is insufficient for the
single-node mode. Smaller local lattices can be also studied by using the 
communication (for instance for thermalization or parameter tuning); 
however, in these cases the communication overhead increases somewhat. 
The performance as a function of the local lattice volume is shown
on Fig.~\ref{myrinet} (we also 
made a direct comparison between our architecture and a Myrinet system
at DESY, Hamburg). Two orders of magnitude smaller local lattice means factor
of $\approx$2 drop in the performance. Thus, for local lattices as small as
$8^3\times 4$ the performance is still acceptable. This indicates clearly the
scalability of this architecture. Note that the performance of a Myrinet based
system is only 5-10\% better than that of the GigE based system.

\section{Conclusion}

We reportde on the status of our PC-based
parallel computer project for lattice QCD (for more details see ref. \cite{FKP02}).
Nearest-neighbor communication 
is implemented in a two-dimensional mesh using 
Gigabit Ethernet cards. 
This architecture presents a good compromise between computation and 
communication. 
The saturated sustained performances on large lattices are around 0.5(0.8)
GFlops/node for staggered (Wilson) fermions, 
which gives a price/performance ratio better than
\$1.0(0.7)/Mflop.\hfill

{\bf Acknowledgment.}
We thank F.~Csikor and Z.~Horv\'ath for their continuous help. 
This work was supported by Hungarian Science Foundation Grants under Contract
Nos. OTKA-T37615/\-T34980/\-T29803/\-M37071/\-OMFB1548/\-OMMU-708.

\end{document}